\documentstyle[12pt]{article}
\def\dif{{\rm d}} \def\deriv{\@ifnextchar[{\@deriv}{\@deriv[]}}
\def\@deriv[#1]#2#3{\mathchoice%
{{\dif^{#1}#2\over\dif{#3}^{#1}}}{{\dif^{#1}#2/\dif{#3}^{#1}}}%
{{\dif^{#1}#2\over\dif{#3}^{#1}}}{{\dif^{#1}#2/\dif{#3}^{#1}}}}

\def\secteqno{\@addtoreset{equation}{section}%
\def\theequation{\thesection.\arabic{equation}}} %
\newcounter{subequation}
\def\thesubequation{\alph{subequation}}
\def\sneqnarray{\stepcounter{equation}\let\@currentlabel=\theequation
\setcounter{subequation}{1} \def\@eqnnum{{\rm
(\theequation.\thesubequation)}}
\global\@eqcnt\z@\tabskip\@centering\let\\=\@eqncr\let\@@eqncr=\@@sneqncr
$$\halign to \displaywidth\bgroup\@eqnsel\hskip\@centering
$\displaystyle\tabskip\z@{##}$&\global\@eqcnt\@ne \hskip 2\arraycolsep
\hfil${##}$\hfil &\global\@eqcnt\tw@ \hskip 2\arraycolsep
$\displaystyle\tabskip\z@{##}$\hfil
\tabskip\@centering&\llap{##}\tabskip\z@\cr}
\def\endsneqnarray{\@@sneqncr\egroup $$\global\@ignoretrue}
\def\@@sneqncr{\let\@tempa\relax \ifcase\@eqcnt \def\@tempa{& & &}\or
\def\@tempa{& &} \else \def\@tempa{&}\fi \@tempa
\if@eqnsw\@eqnnum\stepcounter{subequation}\fi
\global\@eqnswtrue\global\@eqcnt\z@\cr}
\def\ben{\begin{enumerate}} \def\een{\end{enumerate}}
\def\beq{\begin{equation}} \def\eeq{\end{equation}}
\def\bea{\begin{eqnarray}} \def\eea{\end{eqnarray}}
\def\beann{\begin{eqnarray*}} \def\eeann{\end{eqnarray*}}
\def\beasn{\begin{sneqnarray}} \def\eeasn{\end{sneqnarray}}
 
\newcommand{\bref}[1]{(\ref{#1})}

\def\relstack#1#2{\mathrel{\mathop{#1}\limits_{#2}}}
\makeatletter
\def\flE{\begin{picture}(0,0)
   \put( 0.25,    0){\vector( 1, 0){0.50}}
   \@ifstar{\@flE}{\@@flE}}
\def\@flE  #1{\put( 0.5 ,-0.03){\makebox(0,0)[ t]{$#1$}}\end{picture}}
\def\@@flE #1{\put( 0.5 , 0.03){\makebox(0,0)[ b]{$#1$}}\end{picture}}
\def\flNE{\begin{picture}(0,0)
   \put( 0.18, 0.18){\vector( 1, 1){0.64}}
   \@ifstar{\@flNE}{\@@flNE}}
\def\@flNE #1{\put( 0.52, 0.48){\makebox(0,0)[tl]{$#1$}}\end{picture}}
\def\@@flNE#1{\put( 0.48, 0.52){\makebox(0,0)[br]{$#1$}}\end{picture}}
\def\flN{\begin{picture}(0,0)
   \put(    0, 0.20){\vector( 0, 1){0.60}}
   \@ifstar{\@flN}{\@@flN}}
\def\@flN  #1{\put( 0.03, 0.5 ){\makebox(0,0)[ l]{$#1$}}\end{picture}}
\def\@@flN #1{\put(-0.03, 0.5 ){\makebox(0,0)[ r]{$#1$}}\end{picture}}
\def\flNW{\begin{picture}(0,0)
   \put(-0.18, 0.18){\vector(-1, 1){0.64}}
   \@ifstar{\@flNW}{\@@flNW}}
\def\@flNW #1{\put(-0.48, 0.52){\makebox(0,0)[bl]{$#1$}}\end{picture}}
\def\@@flNW#1{\put(-0.52, 0.48){\makebox(0,0)[tr]{$#1$}}\end{picture}}
\def\flW{\begin{picture}(0,0)
   \put(-0.25,    0){\vector(-1, 0){0.50}}
   \@ifstar{\@flW}{\@@flW}}
\def\@flW  #1{\put(-0.5 , 0.03){\makebox(0,0)[ b]{$#1$}}\end{picture}}
\def\@@flW #1{\put(-0.5 ,-0.03){\makebox(0,0)[ t]{$#1$}}\end{picture}}
\def\flSW{\begin{picture}(0,0)
   \put(-0.18,-0.18){\vector(-1,-1){0.64}}
   \@ifstar{\@flSW}{\@@flSW}}
\def\@flSW #1{\put(-0.52,-0.48){\makebox(0,0)[br]{$#1$}}\end{picture}}
\def\@@flSW#1{\put(-0.48,-0.52){\makebox(0,0)[tl]{$#1$}}\end{picture}}
\def\flS{\begin{picture}(0,0)
   \put(    0,-0.2 ){\vector( 0,-1){0.60}}
   \@ifstar{\@flS}{\@@flS}}
\def\@flS  #1{\put(-0.03,-0.5 ){\makebox(0,0)[ r]{$#1$}}\end{picture}}
\def\@@flS #1{\put( 0.03,-0.5 ){\makebox(0,0)[ l]{$#1$}}\end{picture}}
\def\flSE{\begin{picture}(0,0)
   \put( 0.18,-0.18){\vector( 1,-1){0.64}}
   \@ifstar{\@flSE}{\@@flSE}}
\def\@flSE #1{\put( 0.48,-0.52){\makebox(0,0)[tr]{$#1$}}\end{picture}}
\def\@@flSE#1{\put( 0.52,-0.48){\makebox(0,0)[bl]{$#1$}}\end{picture}}
\def\capsa(#1,#2)#3{\put(#1,#2){\makebox(0,0){$#3$}}}
\def\indiag{\@ifnextchar [{\@indiag}{\@indiag[15ex]}}
\def\@indiag[#1](#2,#3){\begingroup
   \setlength{\unitlength}{#1}
   \medskip
   \begin{center}
   \begin{picture}(#2,#3)}
\def\exdiag{\end{picture}
   \end{center}
   \medskip
   \endgroup}
\makeatother

\begin{document}

\title{\bf On Dirac's incomplete analysis of gauge transformations}

\author{Josep M. Pons
\thanks{pons@ecm.ub.es}\\
Departament d'Estructura i Constituents de la Mat\`eria,\\
Facultat de F\'\i sica,
Universitat de Barcelona\\ 
Av.~Diagonal 647, \ 08028 Barcelona\\ Catalonia, Spain
}

\maketitle

\begin{abstract}
Dirac's approach to gauge symmetries is discussed. We follow closely
the steps that led him from his conjecture concerning the
generators of gauge transformations {\it at a given time} ---to be
contrasted with the common view of gauge transformations as maps
from solutions of the equations of motion into other solutions---
to his decision to artificially modify the dynamics, substituting
the extended Hamiltonian (including all first-class constraints)
for the total Hamiltonian (including only the primary first-class
constraints). We show in detail that Dirac's analysis was
incomplete and, in completing it, we prove that the fulfilment of
Dirac's conjecture ---in the ``non-pathological" cases--- does not 
imply any need to modify the dynamics.
We give a couple of simple but significant examples.

\end{abstract}
\newpage


\setcounter{equation}{0}

\section{Introduction }
\label{sec:intro}

It has been more than fifty years since the formal development of
the theory of constrained systems saw the light with the work of
Dirac and Bergmann. By the end of the 1940's, these two
physicists, independently, Peter Gabriel Bergmann with different
collaborators (Bergmann 1949, Bergmann and Brunings 1949,
Anderson and Bergmann 1951) and Paul Adrien
Maurice Dirac (1950), working alone, began the systematic
study of the canonical formalism for what we today call gauge
theories (here including generally covariant theories), also known
---in an almost\footnote{This ``almost'' is to be explained below.}
equivalent terminology--- as constrained systems.

Since these early stages, Bergmann's purpose was nothing other than the
quantization of gravity, whereas for Dirac the purpose was rather the
generalization of Hamiltonian methods, in view also of
quantization, but mostly for special relativistic
theories, and the development of his ideas (1949) on the forms of
relativistic dynamics. Eventually the application to
general relativity made its way indirectly into Dirac's approach
when he considered (1951) the quantization on curved surfaces
 (see also Dirac 1958, 1959,
where general relativity was properly addressed).

The quantization of gravity is an elusive subject that still
remains basically unresolved, because of both technical and
conceptual obstacles. In addition to the fact that general
relativity described gravity as a feature of the geometry of
space-time, there was the problem that, since it was a gauge
theory ---in the form of diffeomorphism invariance---, its Hamiltonian
formulation was unknown, because the standard procedure of
translating the formulation from tangent space ---with a Lagrangian
as a starting point--- to phase space met with some technical
difficulties that had not yet been addressed. Solving these
difficulties was part of the contribution by Bergmann and Dirac.

A Hamiltonian formulation was considered at that time a necessary
step towards quantization: quantization had to proceed through the
correspondence rules ---which were worked out also by Dirac in the
1920's--- that map the classical Poisson brackets of the canonical
variables into commutators of operators. Curiously enough, as a
matter of fact, it was at about the same time as Bergmann's and Dirac's
first contributions to constrained systems, the end of the
1940's, that Richard Feynman (1948) developed the path integral approach
to quantization, which renders the route through the canonical
formalism basically unnecessary\footnote{Note however that the
derivation of the path integral formalism from the canonical
approach is the safest way to guarantee the unitarity of the
S-matrix as well as the correct Feynman rules for some specific
theories (Weinberg 1995).} and restores the Lagrangian function to
its privileged role in defining a Quantum (Field) Theory.

The difficulties that gauge theories pose to their own canonical
formulation were already present in Electromagnetism, but in that
case were somewhat circumvented in a heuristic way by several
methods of fixing the gauge freedom (e.g., Fermi 1932) and by ``ad hoc''
modifications of the Poisson brackets (e.g., Bjorken and Drell 1965)
---thus discovering the Dirac ones, ``avant la lettre''. But a
diffeomorphism invariant theory like general relativity (GR) was
not so easy to tackle. Eventually, a general framework emerged,
applicable to any particular case, that yielded general results on
the canonical formulation of gauge theories. It was mostly Dirac
who gave the final form to the standard formulation of what has been
called thereafter Constrained Systems. His concise but largely
influential Yeshiva ``Lectures on Quantum Mechanics'' (1964)
became the little book from which generations of
theoretical physicists learned the basics of Constrained Systems
and were first acquainted with the key concepts of the
formulation: constraints ---primary, secondary, etc., in a
terminology coined by Bergmann; first-class, second-class, in a
different classification introduced by Dirac---, canonical
Hamiltonian, total Hamiltonian, extended Hamiltonian, arbitrary
functions, gauge transformations, Dirac bracket ---substituting for
the Poisson bracket---, etc. The classical
canonical formulation of GR, the ADM formalism of  
Arnowitt, Deser and Misner (1962), was obtained also 
in the 1960s.

An introductory as well as conceptual overview of Constrained Systems
and Dirac's approach to gauge symmetries can be found in
Earman (2003). There has been a good deal 
of debate concerning generally covariant theories ---like general 
relativity---, where the canonical Hamiltonian is a first class constraint 
(definitions given below)   
participating in the generation of gauge freedom. It has been suggested 
that because of this fact, these theories exhibit no physical dynamics 
in the canonical formalism, in the sense that the dynamics 
seems to be purely gauge. In fact it is not.   
This issue, which is related to the issue of observables for this 
kind of theories, is clarified in Pons and Salisbury (2005). 
Another approach as well as many references can be found in 
Lusanna and Pauri (2003).

\vspace{4mm}

Let us mention at this point that one can find in the literature
other methods of obtaining a canonical formulation for theories
originating from singular Lagrangians. For instance one can adopt
the method of Faddeev and Jackiw (1988), which amounts to
a classical reduction of all the gauge degrees of freedom. The
equivalence of this method with that of Dirac and Bergmann was
shown by Garcia and Pons (1997). Another method, which consists also
in a classical reduction of the gauge degrees of freedom,
considers a quotienting procedure
(Sniatycki 1974, Abraham and Marsden 1978, 
Lee and Wald 1990) to obtain a physical
phase space (endowed with a symplectic form) starting from the
presymplectic form that is defined in the tangent bundle once the singular
Lagrangian is given. It was shown by Pons, Salisbury and Shepley (1999)
that this method is again equivalent to Dirac-Bergmann's.

\vspace{4mm}

The main objective of this paper is to discuss and give a critical
assessment of Dirac's approach to gauge transformations, and his
consequent proposal to modify the dynamics by the use of the
extended Hamiltonian, instead of the one that is obtained from
purely mathematical considerations, the total Hamiltonian. Dirac
only considered gauge transformations \emph{at a given time}, and
this must be contrasted with the most common view of gauge
transformations as {\sl symmetries} that map \emph{entire
solutions} of the dynamics into new solutions, which was 
Bergmann's point of view. We will try to clarify some confusions 
originated from the use of these two different concepts of 
gauge transformation.

 In Section 2 we start with a brief, though almost self-contained,
 introductory Section on Constrained Systems. In Section 3 we
 reproduce verbatim Dirac's own view (1964) on gauge
 transformations whereas in Section 4 we show the limitations of his
 approach and complete it. We make contact with Bergmann's view in
 Section 5. In Section 6 we comment upon the incompleteness of
 Dirac view and its possible explanations. In Section 7 it is shown that
 Dirac's modification of the dynamics has, after all, no
 damaging consequences. Finally we devote the
 last Section to some examples.

\section{Dirac-Bergmann constrained systems in a nutshell}
\label{sec:nut}

Although we are interested in gauge field theories, we will use
mainly the language of mechanics ---that is, of a finite number of
degrees of freedom---, which is sufficient for our purposes. A
quick switch to the field theory language can be achieved by using
DeWitt's (1963) condensed notation. Consider, as our
starting point\footnote{All functions are assumed to be continuous and 
differentiable as many times as the formalism requires.} 
a time-independent first-order Lagrangian $L(q,\,
\dot q)$ defined in configuration-velocity space $TQ$, that is,
the tangent bundle of some configuration manifold $Q$ that we
assume to be of dimension $n$. Gauge theories rely on singular ---as
opposed to regular--- Lagrangians, that is, Lagrangians whose
Hessian matrix with respect to the velocities (where $q$ stands,
in a free index notation, for local coordinates in $Q$),
\begin{equation}
W_{ij}\equiv {\partial^2L\over\partial\dot
q^i\partial\dot q^j},
\label{hess}
\end{equation}
is not invertible.

Two main consequences are drawn from this non-invertibility. First
notice that the Euler-Lagrange equations of motion $[L]_i=0$, with
$$
[L]_i := \alpha_i - W_{ij}\ddot q^j\,,
$$
and
$$
\alpha_i :=
    - {\partial^2L\over\partial\dot q^i\partial q^j}\dot q^j
    + {\partial L\over\partial q^i} \,,
$$
cannot be written in a normal form, that is, isolating on one side
the accelerations, $\ddot q^j = f^j(q,\, \dot q)$. This makes the
usual theorems about the existence and uniqueness of solutions of
ordinary differential equations inapplicable. Consequently, there may be
points in the tangent bundle where there are no solutions passing
through the point, and others where there is more than one
solution. This is in fact our first encounter with constraints and
the phenomenon of gauge freedom. Much more on this will be said
below.

\vspace{4mm}

The second consequence of the Hessian matrix being singular
concerns the construction of the canonical formalism. The Legendre
map from the tangent bundle $TQ$ to the cotangent bundle ---or phase
space--- $T^*Q$ (we use the notation $\hat p(q,\dot q)
 := \frac{\partial L}{\partial\dot q}$),
\begin{eqnarray}
{\mathcal F}\!L :  TQ &\longrightarrow & T^*Q   \\
\qquad  (q,\dot q)&\longrightarrow & (q, p=\hat p)
\end{eqnarray}
is no longer invertible because $\frac{\partial \hat
p}{\partial\dot q}=\frac{\partial L}{\partial\dot q\partial\dot
q}$ is the Hessian matrix. There appears then an issue about the
projectability of structures from the tangent bundle to phase
space: there will be functions defined on $TQ$ that cannot be
translated (projected) to functions on phase space. This feature
of the formalisms propagates in a corresponding way to the
tensor structures, forms, vector fields, etc.

In order to better identify the problem and to obtain the
conditions of projectability, we must be more specific. We will
make a single assumption, which is that the rank of the Hessian
matrix is constant everywhere. If this condition is not satisfied
throughout the whole tangent bundle, we will restrict our
considerations to a region of it, with the same dimensionality,
where this condition holds. So we are assuming that the rank of
the Legendre map ${\mathcal F}\!L$ is constant throughout $TQ$ and
equal to, say, $2n-k$. The image of ${\mathcal F}\!L$ will be locally
defined by the vanishing of $k$ independent functions,
$\phi_\mu(q, p), \mu = 1,2,..,k$. These functions are the {\sl
primary constraints}, and their pullback ${\mathcal
F}\!L^*\phi_\mu$ to the tangent bundle is identically zero:
\begin{equation}({\mathcal F}\!L^*\phi_\mu)(q,\dot q):=\phi_\mu(q, \hat p)
=0,\ \ \forall\, q,\dot q\,. \label{pullback}
\end{equation}

The primary constraints form a generating set of the ideal of
functions that vanish on the image of the Legendre map. With their
help it is easy to obtain a basis of null vectors for the Hessian
matrix. Indeed, applying $\frac{\partial}{\partial\dot q}$ to
\bref{pullback} we get
$$
W_{ij}\left(\frac{\partial \phi_\mu}{\partial
p_j}\right)_{\!|_{p=\hat p}} =0,\ \ \forall\, q,\dot q\,.
$$

With this result in hand, let us consider some geometrical aspects
of the Legendre map. We already know that its image in $T^*Q$ is
given by the primary constraints' surface. A foliation in $TQ$ is
also defined, with each element given as the inverse image of a
point in the primary constraints' surface in $T^*Q$. One can
easily prove that the vector fields tangent to the surfaces of the
foliation are generated by
\begin{equation}
{\bf \Gamma}_\mu =\left(\frac{\partial \phi_\mu}{\partial
p_j}\right)_{\!|_{p=\hat p}}\frac{\partial}{\partial \dot q^j}\,.
\label{gamma}
\end{equation}
The proof goes as follows. Consider two neighboring points in $TQ$
belonging to the same sheet, $(q, \dot q)$ and $(q, \dot q +
\delta \dot q)$ (the configuration coordinates $q$ must be the
same because they are preserved by the Legendre map). Then, using
the definition of the Legendre map, we must have $\hat p(q, \dot
q) =\hat p(q, \dot q + \delta \dot q)$, which implies, expanding
to first order,
$$
\frac{\partial\hat p}{\partial\dot q}\delta \dot q=0\,,
$$
which identifies $\delta \dot q$ as a null vector of the Hessian
matrix (here expressed as $\frac{\partial\hat p}{\partial\dot
q}$). Since we already know a basis for such null vectors,
$\left(\frac{\partial \phi_\mu}{\partial p_j}\right)_{\!|_{p=\hat
p}}, \mu = 1,2,...,k$, it follows that the vector fields ${\bf
\Gamma}_\mu$ form a basis for the vector fields tangent to the
foliation.

\vspace{5mm}

The knowledge of these vector fields is instrumental for
addressing the issue of the projectability of structures. Consider
a real-valued function $f^L : TQ \longrightarrow \mathbf{R}$. It will
---locally--- define a function $f^H : T^*Q \longrightarrow \mathbf{R}$ iff
it is constant on the sheets of the foliation, that is, when
\begin{equation}
{\bf  \Gamma}_\mu f^L =0,\ \ \mu = 1,2,...,k\,.\label{proj-cond}
\end{equation}

Equation \bref{proj-cond} is the projectability condition we were
looking for. We express it in the following way:
$${\bf  \Gamma}_\mu f^L =0,\ \ \mu = 1,2,...,k \Leftrightarrow
{\rm there \ \, exists}\ \, f^H {\rm such \ \, that} \ \,{\mathcal
F}\!L^*f^H =f^L \,.
$$

\subsection{The canonical Hamiltonian}

A basic ingredient of the canonical formalism is the Hamiltonian
function. In the case of a regular theory (that is, with a
non-singular Hessian matrix) it defines, by use of the Poisson
bracket, the vector field that generates the time evolution ---the
dynamics--- in phase space. The Hamiltonian is given in that case
as the projection to phase space of the Lagrangian energy $E =
\frac{\partial L}{\partial \dot q } - L$.

This procedure to define the Hamiltonian will still work in the
singular case if the energy satisfies the conditions of
projectability \bref{proj-cond}. Indeed we can readily check that
${\bf  \Gamma}_\mu E = 0$, so we have a {\sl canonical
Hamiltonian} $H_c$, defined as a function on phase space whose
pullback is the Lagrangian energy, ${\mathcal F}\!L^*H_c =E$.  It
was Dirac that first realized in the general setting of
constrained systems that a Hamiltonian always existed.

There is a slight difference, though, from the regular case, for
now there is an ambiguity in the definition of $H_c$. In fact,
since ${\mathcal F}\!L^*\phi_\mu =0$, many candidates for canonical
Hamiltonians are available, once we are given one. In fact, $H_c +
v^\mu \phi_\mu$ ---with $v^\mu(q,\dot q;t)$ arbitrary functions and
with summation convention for $\mu$--- is as good as $H_c$ as a
canonical Hamiltonian. This ``slight difference'' is bound to have
profound consequences: it is the door to gauge freedom.

\subsection{Dynamics for constrained systems} The Hamiltonian ---with
arbitrary func\-tions--- $H_c + v^\mu \phi_\mu$ was called by Dirac
the total Hamiltonian, although today it is usually referred to as
the Dirac Hamiltonian $H_D$.

In the regular case, once the Hamiltonian is given, the equations
of motion in phase space are deterministically formulated as
\begin{equation}
\dot q = \{q, \, H_c  \}, \ \ \dot p = \{p, \, H_c  \}\,.
\end{equation}
So in the singular case we could try, taking into account the
non-uniqueness of the canonical Hamiltonian,
\begin{eqnarray}
\dot q &= & \{q, \, H_D  \}=\{q, \, H_c  \} +
v^\mu \{q, \, \phi_\mu\} \,, \nonumber \\
\dot p &= &  \{q, \, H_D  \}=\{p, \, H_c  \} +
v^\mu \{p, \, \phi_\mu\}\,,\nonumber \\
0& =&\phi_\mu(q,p)  \,. \label{jump}
\end{eqnarray}
Of course, as of now, this formulation \bref{jump} is just a
reasonable guess. But it turns out that it is {\sl correct} in a
very precise sense, as already shown by Dirac (1950). To be
a bit more precise, it was proven by Batlle, Gomis, Pons and Roman (1986)
that the
equations \bref{jump} are equivalent to the Euler-Lagrange
equations $[L]_i=0$. This is to say that if $(q(t), p(t))$ is a
trajectory in phase space satisfying \bref{jump}, then $q(t)$ is a
solution of the Euler-Lagrange equations. And vice-versa, if
$q(t)$ is a solution of the Euler-Lagrange equations, then the
definition $p(t) := \hat p(q(t), \frac{d q}{d t})$ makes $(q(t),
p(t))$ a solution of \bref{jump}. Note that although the arbitrary
functions $v^\mu$ may depend on the time and the phase space
variables, on a given solution $(q(t), p(t))$ they become just
functions of the time variable. That is, assuming that the
functions $v^\mu$ are simply arbitrary functions of time is
sufficient in order to describe all solutions to the system
\bref{jump}.

So we have succeeded in obtaining a Hamiltonian formulation for a
theory defined through a singular Lagrangian. It is worth noticing
that the Hamiltonian equations of motion have two parts, a
differential one, corresponding to the first two lines in
\bref{jump}, and an algebraic one, which is the third line ---the
primary constraints. Both types of equations are coupled in the
sense that the constraints may impose severe restrictions on the
solutions of the differential equations ---or even may prevent them
from existing. Dirac devised a clever way to disentangle the
algebraic and differential components of \bref{jump}, which we are
going to summarise in the next two subsections. Let is emphasize,
however, that the formulation of the dynamics is already complete
in \bref{jump}, and that the developments below are just
convenient elements for dealing with equations of the type
\bref{jump}.
\subsection{Dirac's classification of constraints}

The Hamiltonian time evolution vector field, derived from the
differential part in \bref{jump} is given by
\begin{equation} {\bf X_H}:=\frac{\partial }{ \partial t} +
\{-,\,H_c \} + v^\mu \{-,\, \phi_\mu \}, \label{xh}
\end{equation} where $v^{\mu}$ are arbitrary functions of
time and we have introduced $\frac{\partial }{ \partial t}$ to
account for possible explicit dependences on time.

Let us now examine the marriage between the algebraic and
differential parts in \bref{jump}. First, we require the
preservation in time of the primary constraints, that is, $\bf X_H
\phi_{\mu} = 0$ on any trajectory solution of \bref{jump}. These
are tangency conditions that may lead to new constraints and to
the determination of some of the functions $v^\mu$. Here enters
Dirac's clever idea of splitting the primary constraints in two
types: those that are first-class, $\phi_{\mu_0}$, and the rest,
called second class, $\phi_{\mu_1}$. They are defined respectively
by
\begin{equation}
\{ \phi_{\mu_0},\,\phi_{\mu} \} \relstack{=}{pc} 0\,, \qquad {\rm
and} \qquad\det |\{ \phi_{\mu_1},\,\phi_{\nu_1} \}|
\relstack{\not=}{pc} 0\,, \label{fc-sc}
\end{equation} where $\{-,\,- \}$ is as before the Poisson Bracket
and $pc$ stands for a generic linear combination of the {\sl
primary} constraints. The subscript $pc$ under the sign of
equality (or inequality) means that such equality (or inequality)
holds for all the points in phase space that lie on the primary
constraints' surface. Let us mention the technical point that
sometimes the inequality above does not hold for every point; this
fact rises the issue that some constraints initially classified as
second-class may eventually become first-class when new
constraints appear in the formalism. We will not consider such a
situation and will assume henceforth that the determinant in
(\ref{fc-sc}) will be different from zero everywhere on the
surface of primary constraints.

Note that the concept of a function being first-class is not
restricted to functions representing constraints. In fact, we say
that a function $f$ is first-class with respect to a given set of
constraints if its Poisson bracket with these constraints vanishes
in the constraints' surface.

\subsection{Refining the dynamics}

The requirement of the tangency of $\bf X_H$ to the second class
constraints fixes some arbitrariness in the Hamiltonian dynamics.
The arbitrary functions $v^{\nu_1}$ ---where $\nu_1$ runs over the
indices of the secondary constraints--- become determined as
canonical functions $v_c^{\nu_1}$ through
\begin{equation} 0 = {\bf X_H}\phi_{\mu_1}=\{\phi_{\mu_1},\,H_c \} +
v_c^{\nu_1}
\{\phi_{\mu_1},\, \phi_{\nu_1} \}\,, \label{determ}
\end{equation}
which yields
$$ v_c^{\mu_1} = - M^{{\mu}_1{\nu}_1} \{\phi_{\nu_1},\,H_c
\}\,, $$
where $M^{{\mu}_1{\nu}_1}$ is the matrix inverse of the
Poisson bracket matrix of the primary second-class constraints,
$\{ \phi_{\mu_1},\,\phi_{\nu_1} \}$.

Substituting $v^{\nu_1}_c$ for $v^{\nu_1}$ in \bref{xh} gives a
more refined expression for the dynamics:
\begin{equation}
{\bf X^1_H} :={\partial \over \partial t} + \{-,\,H_c \}^* +
v^{\mu_0} \{-,\,\phi_{\mu_0} \}\,, \label{xh1}
\end{equation}
where a new structure, the Dirac bracket, has been introduced, at
this level of the primary constraints, by the definition
$$
\{A,\,B\}^* := \{A,\,B\} - \{A,\,\phi_{\mu_1}\}M^{{\mu_1}{\nu_1}}
\{\phi_{\nu_1},\,B\} \,,
$$

Next we must require the dynamics to preserve the primary
first-class constraints $\phi_{\mu_0}$. The definition of the
first class property in \bref{fc-sc} makes irrelevant the choice
between the initial form of the dynamics and the refined form. For
with either choice this requirement ends up as the condition
$$\{\phi_{\nu_0},\,H_c \} = 0\,,
$$
on any solution of \bref{jump}. If, for some $\nu_0$, the bracket
$\{\phi_{\nu_0},\,H_c \}$ already gives zero on the primary
constraints' surface, nothing new needs to be done, but if
$\phi^1_{\nu_0}:= \{\phi_{\nu_0},\,H_c \}$ is different from zero
on that surface, it means that we have found {\sl new constraints}
that further restrict the region where solutions to \bref{jump}
may exist. These $\phi^1_{\nu_0}$ (for the appropriate
${\nu_0}$'s) are called the {\sl secondary} constraints
(Anderson and Bergmann 1951).

Note that the evolutionary operator \bref{xh1} can be
alternatively expressed (taking into account the fulfillment of
the primary constraints) as
\begin{equation} {\bf X^1_H} :={\partial
\over \partial t} + \{-,\,H_c^* \} + v^{\mu_0} \{-,\,\phi_{\mu_0}
\}\,, \label{xh1*}
\end{equation}
with $H_c^*$ itself a new canonical hamiltonian defined by
\begin{equation}
H_c^* := H_c
-\{H_c,\,\phi_{\mu_1}\}M^{{\mu_1}{\nu_1}}\phi_{\nu_1},
\label{h-star}
\end{equation}
thus making the use of the Dirac bracket unnecessary.

 Summing up,
an initial analysis of the internal consistency of the system
\bref{jump} has led us to the equivalent system
\begin{eqnarray}
\dot q &= & \{q, \, H_c^*  \} +
v^{\mu_0} \{q, \, \phi_{\mu_0}\} \,, \nonumber\\
\dot p &= & \{p, \, H_c^*  \} +
v^{\mu_0} \{p, \, \phi_{\mu_0}\}\,,\nonumber\\
0& =&\phi_\mu(q,p)  \nonumber\\
0& =&\phi^1_{\nu_0}(q,p) \,, \label{jump2}
\end{eqnarray}
which is a first step in our endeavour to decouple the differential
and the algebraic sides in \bref{jump}. In the language of the
trade, we have undertaken the first step in the Dirac {\sl constraint
algorithm}.

Now the way is paved for the next steps to be taken. If new ---i.e.
secondary--- constraints have been introduced in the first step, we
must ask again for the tangency of the new evolution operator,
\bref{xh1} or \bref{xh1*}, to them. This requirement may bring
some of the formerly primary first-class constraints into the
second-class category (thus producing the determination of some of
the remaining arbitrary functions) and, again, may give new
---tertiary--- constraints. We will not dwell on the details, easily
reconstructed, but just mention that the application of the
algorithm ends when we reach a final constraint surface to which
the final form of the time evolution vector field is already
tangent, so that no more constraints appear and no more arbitrary
functions get determined by consistency requirements. This final
evolution vector field will be written as
\begin{equation} {\bf X^F_H} :={\partial
\over \partial t} + \{-,\,H' \} + v^{\mu'} \{-,\,\phi_{\mu'} \}\,,
\label{xhf}
\end{equation}
with $H'$ the final, first-class, Hamiltonian\footnote{Note that
$H'$ is a specific choice of a canonical Hamiltonian.}, and
$\phi_{\mu'}$ the final primary first-class constraints.

So at this final stage a certain set of constraints, primary,
secondary, tertiary, etc., will restrict the region of phase space
where a solution can exist. Let us denote these generic
constraints as $\phi_A$, for some index $A$ that will run through
the whole set of primary, secondary, tertiary, etc., constraints.
So the final picture of the dynamics will be expressed with a
system of equations equivalent to \bref{jump},
\begin{eqnarray}
\dot q &= & \{q, \, H'  \} + v^{\mu'} \{q, \, \phi_{\mu'}\} \,, \nonumber\\
\dot p &= & \{p, \, H'  \} + v^{\mu'} \{p, \, \phi_{\mu'}\}\,,\nonumber\\
0& =&\phi_A(q,p)  \,, \label{jump-f}
\end{eqnarray}
with $v^{\mu'}$ the arbitrary functions associated with the final
primary first-class constraints $\phi_{\mu'}$.

Note the crucial difference between the initial equations
\bref{jump} and the final ones \bref{jump-f}. Since $H'$ and
$\phi_{\mu'}$ are first-class with respect to the whole set of
constraints $\phi_A$, we only need to choose the initial
conditions ---at, say, $t=0$--- $(q(0), p(0))$ in such a way that the
constraints are satisfied. Then, for whatever arbitrary functions
we may use for $v^{\mu'}$, the solution of the differential
equations in the first two lines in \bref{jump-f} will always
satisfy the constraints. The differential and the algebraic sides
in \bref{jump-f} are now completely disentangled.

The presence of arbitrary functions in the final form of the
dynamics \bref{jump-f} signals the existence of gauge freedom,
which will be the subject of the next section. Note that there may
exist constrained systems (that is, systems described by singular
Lagrangians) that do not exhibit any gauge freedom, because all
constraints eventually become second class. That is why the
phrases of ``gauge theories'' and ``constrained systems'' are not
entirely equivalent.

Finally, for further use, let us mention the notation invented by
Dirac for the concepts of weak ($\approx$) and strong ($\equiv$)
equalities, with respect to a set of constraints that we denote
generically by $\phi$. A function $f$ is said to be \emph{weakly
equal} to zero,
$$f \approx 0\,,$$
if it vanishes on the surface
defined by the constraints, $ f \relstack{=}{\phi=0}0$. A function
$f$ is said to be \emph{strongly equal} to zero,
$$f \equiv 0\,,$$
if both $f$ and its differential ---that is, its partial
derivatives $\frac{\partial f}{\partial q},\ \frac{\partial
f}{\partial p}$--- vanish on the surface defined by the
constraints, $ f \relstack{=}{\phi=0}0$, $ df
\relstack{=}{\phi=0}0$.
\section{Gauge freedom: Dirac's view}

As we said, when the final equations \bref{jump-f} exhibit
arbitrary functions in the dynamics, the phenomenon of gauge
freedom is present in our formulation, and there will exist
\emph{gauge transformations} connecting different solutions of
\bref{jump-f} that share the same initial conditions. From the
mathematical point of view, the dynamics is no longer
deterministic.

Now we will reproduce in literal terms Dirac's analysis of gauge
transformations. Let us say at the outset that, as the title of
this paper indicates, we shall eventually find this analysis
incomplete; but in this Section we will accurately reproduce
Dirac's view in his own words. Our comments will be reserved for
the next section. The source here will be exclusively the little
book (1964), which was written when the theory of
constrained systems was settled enough, and which probably
represents Dirac's mature perspective on the subject.


\vspace{4mm}

 \underline{\bf Dirac, verbatim:}

\vspace{4mm}

\begin{quote}
Let us try to get a physical understanding of the situation
where we start with given initial variables and get a solution of
the equations of motion containing arbitrary functions. The
initial variables which we need are the $q$'s and the $p$'s. We
don't need to be given initial values for the coefficients $v$.
These initial conditions describe what physicists would call the
\emph{initial physical state} of the system. The physical state is
determined only by the $q$'s and the $p$'s and not by the
coefficients $v$.

Now the initial state must determine the state at later times. But
the $q$'s and the $p$'s at later times are not uniquely determined
by the initial state because we have the arbitrary functions $v$
coming in. That means that the state does not uniquely determine a
set of $q$'s and $p$'s, even though a set of $q$'s and $p$'s
uniquely determines a state. There must be several choices of
$q$'s and $p$'s which correspond to the same state. So we have the
problem of looking for all the sets of $q$'s and $p$'s that
correspond to one particular physical state.

All those values for the $q$'s and $p$'s at a certain time which
can evolve from one initial state must correspond to the same
physical state at that time. Let us take particular initial values
for the $q$'s and the $p$'s at time $t=0$, and consider what the
$q$'s and the $p$'s are after a short time interval $\delta t$.
For a general dynamical variable $g$, with initial value $g_0$,
its value at time $\delta t$ is
\begin{eqnarray}
g(\delta t) &=& g_0 + \delta t\,\dot g \\ \nonumber &=& g_0 +
\delta t\,\{g,\, H_T\} \\ \nonumber &=& g_0 + \delta t (\{g,\,
H'\} + v^a\{g,\, \phi_a\})\,.
\end{eqnarray}

The coefficients $v$ are completely arbitrary and at our disposal.
Suppose we take different values, $v'$, for these coefficients.
That would give a different $g(\delta t)$, the difference being
\begin{equation}
\Delta g(\delta t)=\delta t (v'^a - v^{a})\{g,\, \phi_a\}\,.
\end{equation}

We may write this as
\begin{equation}
\Delta g(\delta t)=\epsilon^a\{g,\, \phi_a\}\,,
\label{rule}\end{equation} where
\begin{equation}
\epsilon^a=\delta t (v'^a - v^{a})\label{rule2}
\end{equation}
is a small arbitrary number, small because of the coefficient
$\delta t$ and arbitrary because the $v$'s and $v'$'s are
arbitrary. We can change all our Hamiltonian variables in
accordance with the rule \bref{rule} and the new Hamiltonian
variables will describe the same state. This change in the
Hamiltonian variables consists in applying an infinitesimal
contact transformation with a generating function $\epsilon^a
\phi_a$. We come to the conclusion that the $\phi_a$'s, which
appeared in the theory in the first place as the primary
first-class constraints, have this meaning: \emph{as generating
functions of infinitesimal contact transformations, they lead to
changes in the $q$'s and the $p$'s that do not affect the physical
state}.\ (Dirac 1964, p 20-21)

\end{quote}

\begin{center}---------------
\end{center}

 Dirac next shows in extreme detail that applying,
after $\epsilon^a \phi_a$, a second contact transformation
$\gamma^a \phi_a$, reversing the order and subtracting, and using the
Jacobi identity for the Poisson brackets, one gets
\begin{equation}
\Delta g=\epsilon^a\gamma^b\{g,\, \{\phi_a,\, \phi_b\}\}\,,
\label{}\end{equation} and then he infers that: 

\vspace{3mm}
\begin{quote}
This $\Delta g$ must also correspond to a change in the $q$'s
and the $p$'s which does not involve any change in the physical
state, because it is made up by processes which individually don't
involve any change in the physical state. Thus we see that we can
use
\begin{equation}
\{\phi_a,\, \phi_b\} \label{fc-bracket}
\end{equation}
as a generating function of an infinitesimal contact
transformation and it will still cause no change in the physical
state.

Now the $\phi_a$ are first-class: their Poisson brackets are
weakly zero, and therefore strongly equal to some linear function
of the $\phi$'s. This linear function of the $\phi$'s must be
first-class because of the theorem I proved a little while back,
that the Poisson bracket of two first-class quantities is
first-class. So we see that the transformations which we get this
way, corresponding to no change in the physical state, are
transformations for which the generating function is a first-class
constraint. The only way these transformations are more general
than the ones we had before is that the generating functions which
we had before are restricted to be first-class primary
constraints. Those that we get now could be first-class secondary
constraints. The result of this calculation is to show that we
might have a first-class secondary constraint as a generating
function of an infinitesimal contact transformation which leads to
a change in the $q$'s and the $p$'s without changing the state.

For the sake of completeness, there is a little bit of further
work one ought to do which shows that a Poisson bracket
$\{H',\,\phi_a\}$ of the first-class hamiltonian with a
first-class $\phi$ is again a linear function of first-class
constraints. This can also be shown to be a possible generator for
infinitesimal contact transformation which do not change the
state.

The final result is that those transformations of the dynamical
variables which do not change physical states are infinitesimal
contact transformations in which the generating function is a
primary first-class constraint or possibly a secondary first-class
constraint. A good many of the secondary first-class constraints
turn up by the process \bref{fc-bracket} or as $\{H',\,\phi_a\}$.
I think it may be that all the first-class secondary constraints
should be included among the transformations which don't change
the physical state, but I haven't been able to prove it. Also, I
haven't found any example for which there exist first-class
secondary constraints which do generate a change in the physical
state.\ (Dirac 1964, p 22-23-24)
\end{quote}
\begin{center}---------------
\end{center}
\vspace{3mm}
\begin{quote}
We were led to the idea that there are certain changes in the
$p$'s and the $q$'s that do not correspond to a change of state,
and which have as generators first-class secondary constraints.
That suggests that one should generalize the equations of motion
in order to allow as variations of a dynamical variable $g$ with
the time not only any variation given by
$$
\dot g = \{g,\, H_T\}\,,
$$
but also any variation which does not correspond to a change of
state. So we should consider a more general equation of motion
$$
\dot g = \{g,\, H_E\}
$$
with an extended Hamiltonian $H_E$, consisting of the previous
Hamiltonian $H_T$, plus all those generators that do not change
the state, with arbitrary coefficients:
$$
 H_E = H_T + v'^{a'}\phi_{a'}.
$$
Those generators $\phi_{a'}$, which are not included already in
$H_T$, will be the first-class secondary constraints. The presence
of these further terms in the Hamiltonian will give further
changes in $g$, but these further changes in $g$ do not correspond
to any change of state and so they should certainly be included,
even though we did not arrive at these further changes of $g$ by
direct work from the Lagrangian."\ (Dirac 1964, p 25)
\end{quote}
\begin{center}---------------
\end{center}
\vspace{3mm}
\begin{quote}
You notice that when we have passed over to the quantum theory,
the distinction between primary constraints and secondary
constraints ceases to be of any importance....Once we have gone
over to the Hamiltonian formalism we can really forget about the
distinction between primary constraints and secondary constraints.
The distinction between first-class and second-class constraints
is very important."\ (Dirac 1964, p 43)
\end{quote}

\section{Gauge freedom revisited: the incompleteness of Dirac's view}

The limitation of Dirac's analysis is that he only examined the
gauge transformations in an infinitesimal neighborhood of the
initial conditions. This is shown by his using an infinitesimal
parameter, that he took as $\delta t$, and an arbitrary function,
which he took as the difference $v^a - v'^{a}$. Instead, we shall
proceed to examine gauge transformations at any value of the
parameter $t$, that is, gauge transformations for the entire
trajectory. The infinitesimal parameter will no longer be $\delta
t$, but a new $\delta s$ that describes an infinitesimal motion
that maps a trajectory into another in such a way that points are
mapped into points corresponding to the {\sl same time}. This new
$\delta s$, times an arbitrary function $f^a$, will describe the
difference $v'^a - v^{a}$, which now is taken to be infinitesimal. So
the infinitesimal parameter and the arbitrary function appear
together in
$$\epsilon^a := v'^a - v^{a} = \delta v^a = \delta s \,f^a.$$ The two
total Hamiltonians  $H' + v^a \phi_a$ and $H' + v'^a \phi_a$,
differ by an infinitesimal arbitrary function \emph{for any value
of the time parameter $t$}. This proposal goes beyond the scope of
Dirac's work, which was circumscribed to an infinitesimal
neighborhood of the canonical variables describing the trajectory
at time $t=0$.

We will see that when we complete the work by Dirac, the
generators of gauge transformations (mapping solutions into
solutions) will be characterized\footnote{We present a slightly
modified derivation from that in Gracia and Pons (1988), see also 
Banerjee, Rothe and Rothe (1999) for a parallel derivation.} 
by nice mathematical expressions which, when restricted to the
infinitesimal region examined by Dirac, that is, around the
initial conditions, will reproduce his results.

\vspace{6mm}

Consider\footnote{We keep using Dirac's notation: $H'$ is a
first-class canonical Hamiltonian and $\phi_a$ are the final
primary first-class constraints.} a dynamical trajectory generated
by the total Hamiltonian $H_T = H' + v^a \phi_a$ out of some
initial conditions at $t=0$. To emphasise the role of the
arbitrary functions
$v^a$, let us use the notation $g_v(t)$ for it. This trajectory
satisfies the equations of motion
$$
\dot g_v(t) =\{g,\, H'\}_{{}_{\!g_v(t)}} + v^a\{g,\,
\phi_a\}_{{}_{\!g_v(t)}}\,.
$$
An infinitesimally close trajectory, sharing the same initial
conditions and generated by $ H' + v'^a \phi_a$, with $v'â = v^a +
\delta v^a$, will be denoted by $g_{v'}$. Let us define the
variation $\Delta g = g_{v'}-g_{v}$, which is an equal-time
variation, that is, $\Delta g(t) = g_{v'}(t)-g_{v}(t)$. Because of
that, this variation commutes with the time derivative,
\begin{equation}[\Delta,\, \frac{d}{d\,t}]=0\,.
\label{commut}
\end{equation}
We shall make extensive use of this fact in the following.

First, notice that $\Delta g(t)$ can be conceived as the result of
a chain of canonical transformations: bringing $g_v(t)$ down to
$g_v(0) = g_{v'}(0)$ through the time evolution generated by $ H'
+ v^a \phi_a$ (going backwards in time) and then up to $g_{v'}(t)$
using $ H' + v'^a \phi_a$. Therefore $\Delta g(t)$ is an
infinitesimal canonical transformation that can be written as
\begin{equation}
\Delta g(t)=\{g,\, G(t)\}_{{}_{\!g_{\!v}}}\label{delta-g}
\end{equation}
for some function $G(t)$\footnote{Observe that $G(t)$ is hiding an
infinitesimal factor $\delta s$ which for the sake of simplicity
we do not make explicit.} in phase space, that is, a function
$G(q,p;t)$. Note that a change of the coefficient functions $v^a$
will also produce changes in $G(t)$. Since the vector field
$\{-,\, G(t)\}$ generates a map from solutions into solutions, it
must preserve the constraints of the theory, therefore $G(t)$ is
first-class function.

Now let us use \bref{commut} to compute $\Delta \dot g$ in two
different ways (we use $v'^a= v^a+\delta v^a$ and keep terms up to
first order in $\delta v^a$).

 \vspace{5mm}

\textbf{First way}
\begin{eqnarray} \nonumber
\Delta \dot g &=&\!\dot g_{\!v'}-\dot g_{\!v} = \{g,\,
H'\}_{{}_{\!g_{\!v+\delta v}}} + (v^a + \delta v^a)\{g,\,
\phi_a\}_{{}_{\!g_{\!v+\delta v}}} -\{g,\,
H'\}_{{}_{\!g_{\!v}}}- v^a \{g,\, \phi_a\}_{{}_{\!g_{\!v}}}\\
\nonumber &=&\Big(\{g,\, H'\}_{{}_{\!g_{\!v+\delta v}}}-\{g,\,
H'\}_{{}_{\!g_{\!v}}}\Big) +\Big(v^a \{g,\,
\phi_a\}_{{}_{\!g_{\!v+\delta v}}}- v^a \{g,\,
\phi_a\}_{{}_{\!g_{\!v}}}\Big)+ \delta v^a \{g,\,
\phi_a\}_{{}_{\!g_{\!v}}}
\\
\nonumber &=&\big(\frac{\partial\{g,\, H'\}}{\partial
g}\big)_{{}_{\!g_{\!v}}} \Delta  g + v^a \big(\frac{\partial\{g,\,
\phi_a\}}{\partial g}\big)_{{}_{\!g_{\!v}}}\Delta  g + \delta v^a
\{g,\,
\phi_a\}_{{}_{\!g_{\!v}}}\\
\nonumber &=& \big(\frac{\partial\{g,\, H'\}}{\partial
g}\big)_{{}_{\!g_{\!v}}} \{g,\, G\}_{{}_{\!g_{\!v}}} + v^a
\big(\frac{\partial\{g,\, \phi_a\}}{\partial
g}\big)_{{}_{\!g_{\!v}}}\{g,\, G\}_{{}_{\!g_{\!v}}} + \delta v^a
\{g,\,
\phi_a\}_{{}_{\!g_{\!v}}}\\
\nonumber &=& \Big(\{ \{g,\, H'\}\,G\} + v^a \{ \{g,\,
\phi_a\}\,G\} + \delta v^a \{g,\, \phi_a\}\Big)_{{}_{\!g_{\!v}}}\\
\nonumber &=& \Big(\{ \{g,\, H_T\}\,G\} + \delta v^a \{g,\,
\phi_a\}\Big)_{{}_{\!g_{\!v}}}\,.
\end{eqnarray}

\textbf{Second way}
\begin{eqnarray} \nonumber
\Delta \dot g &=& \frac{d}{d\,t} \Delta  g =\frac{d}{d\,t} \{g,\,
G\}_{{}_{\!g_{\!v}}} \\
\nonumber &=& \frac{\partial}{\partial \,t}\{g,\,
G\}_{{}_{\!g_{\!v}}} +\{\{g,\, G\},\, H_T\}_{{}_{\!g_{\!v}}}\\
\nonumber &=& \Big(\{g,\,\frac{\partial\, G}{\partial \,t}\}
+\{\{g,\, G\},\, H_T\}\Big)_{{}_{\!g_{\!v}}}\,.
\end{eqnarray}

Now, comparing these two expressions for $\Delta \dot g$ and using
the Jacobi identities for the Poisson brackets leads to
$$
\{g,\,\frac{\partial\, G}{\partial \,t} + \{ G,\, H_T\} - \delta
v^a \phi_a \}_{{}_{\!g_{\!v}}} =0\,.
$$

At any time, the point $g_{v}(t)$ in phase space, for a generic
dynamical trajectory $g_{v}$, can be any point on the surface of
constraints. Thus, freed from a specific trajectory, the
contents of the previous expression is just the weak equality
$$
\{g,\,\frac{\partial\, G}{\partial \,t} + \{ G,\, H_T\} - \delta
v^a\, \phi_a \} \approx 0\,.
$$
The variable $g$ represents any canonical variable, therefore the
last expression is equivalent to the strong equality
$$\frac{\partial\,
G}{\partial \,t} + \{ G,\, H_T\} - \delta v^a \,\phi_a  \equiv
f(t)\,,
$$
for some function $f$ that depends exclusively on the time
parameter $t$. A trivial redefinition of $G$, $$G(t) \rightarrow
G(t) - \int^t d\tau \, f(\tau)$$ makes this function disappear
without affecting \bref{delta-g}. We obtain, with the redefined
$G$,
\begin{equation}
\frac{\partial\, G}{\partial \,t} + \{ G,\, H_T\} - \delta v^a\,
\phi_a  \equiv 0\,.\label{final1}
\end{equation}
Recall that the $\delta v^a$'s are arbitrary infinitesimal
functions, and the $\phi_a$'s are the primary first-class
constraints (\textbf{pfcc}). Then \bref{final1} can be
alternatively written with no mention to the $\delta v^a$'s,
\begin{equation} \frac{\partial\,
G}{\partial \,t} + \{ G,\, H_T\}  \equiv {\rm \bf pfcc}
\,.\label{final2}
\end{equation}
Finally, recalling that $H_T = H' + v^a \phi_a$ and that the
functions $v^a$ are arbitrary as well, we get the three conditions
for $G(t)$ to be a canonical generator of infinitesimal gauge
transformations:
\begin{equation} G(t) {\rm \ \, is \ \, a \ \,first \
\,class\ \, function},\label{final3}\end{equation}

\begin{equation} \frac{\partial\,
G}{\partial \,t} + \{ G,\, H'\}  \equiv {\rm \bf pfcc}
\,,\label{final4}
\end{equation}
\begin{equation}  \{ G,\, \phi_a\}  \equiv {\rm \bf pfcc}
\,.\label{final5}
\end{equation}

Note that putting {\bf pc} (primary constraints) instead of {\bf
pfcc} in \bref{final4} and \bref{final5} would have been
sufficient because the first class condition in these equations is
already guaranteed by \bref{final3} taken together with the fact
that $H'$ and $\phi_a$ are first class.

 Let us briefly comment on our result. We have found that,
in addition to being first-class, $G$ is a constant of motion for
the dynamics generated by $H_T$, for any values of the
arbitrary functions $v^a$. This is just the meaning of
\bref{final4} and \bref{final5}. It is a constant of motion of a
very specific type, as is seen directly in \bref{final2}. One must
notice that, in contrast with the case of regular theories,
Dirac-Bergmann constrained systems have different types of
constants of motion, according to the status of what appears in
the right side of \bref{final2}. For instance, if instead of the
strong equality to a linear combination of primary first-class
constraints, we had a strong or weak equality to \emph{any}
constraint, we still would have a constant of motion, but if
\bref{final2} is not satisfied, it will not generate a gauge
transformation.

But in fact our result goes beyond the consideration of gauge
transformations. We have just found the conditions for $G(t)$ in
\bref{delta-g} to generate a symmetry, either rigid or gauge.
\emph{Any object $G$ satisfying the three conditions above is a
canonical generator of a symmetry of the dynamics, that maps
solutions into solutions}. These symmetries may depend on
arbitrary functions (more on this below) and then will be called
gauge symmetries (or gauge transformations). If they do not depend
on arbitrary functions they will be called rigid symmetries.
\emph{What we have found in the three conditions
\bref{final3}, \bref{final4}, \bref{final5} is the characterization
of the generators of symmetries in phase space that are canonical
transformations.}

It is worth noticing that conditions
\bref{final3}, \bref{final4}, \bref{final5} come very close to
saying that $G$ is a Noether conserved quantity, thus generating a
Noether symmetry through \bref{delta-g}. Indeed this would have
been the case if our theory had been defined by a regular
Lagrangian, and not by a singular one. But in gauge systems,
Noether theory has some specific features. Let us just mention
that the characterisation of a conserved quantity that generates a
Noether symmetry projectable from tangent space to phase
space\footnote{In gauge theories, there may be Noether symmetries
in the tangent bundle that are \emph{not} projectable to phase
space. This case has been discussed in Garcia and Pons (2000), see
also Garcia and Pons (2001), Gracia and Pons (2000), 
Gracia and Pons (2001).} is given in Batlle, Gomis, Gracia and Pons (1989)
by the following conditions

\begin{equation} \frac{\partial\,
G}{\partial \,t} + \{ G,\, H_c\}  = {\rm \bf pc}
\,,\label{final4bis}
\end{equation}
\begin{equation}  \{ G,\, {\rm \bf pc} \}  = {\rm \bf pc}
\,.\label{final5bis}
\end{equation}

Note that the fulfillment of \bref{final4bis}, \bref{final5bis}
makes $G$ first class. Equations
\bref{final4bis}, \bref{final5bis} are more restrictive than
\bref{final3}, \bref{final4}, \bref{final5} in three ways. (a) The
strong equality there is replaced here by an ordinary equality.
(b) In \bref{final5bis} the Hamiltonian is the canonical one,
$H_c$, which, unlike $H'$, is not necessarily first-class. (c) In
the Poisson bracket in \bref{final5bis} all primary constraints
appear in the lhs, and not only those that are first-class.

\section{Bergmann's version of gauge transformations}

Probably inspired by the examples of electromagnetism, where the
gauge transformation of the gauge potential is $$\delta A_{\mu} =
\{A_{\mu}, \, G\} = \partial_\mu{\Lambda}\,,$$ for an arbitrary
function $\Lambda$, and general relativity, where the gauge
transformations (diffeomorphisms) for the metric field read
$$\delta g_{\mu\nu} = \epsilon^\rho \partial_\rho g_{\mu\nu} +
g_{\mu\rho} \partial_\nu \epsilon^\rho + g_{\rho\nu} \partial_\mu
\epsilon^\rho\,,$$ (for some arbitrary functions $\epsilon^\rho$, 
components of an arbitrary vector field),
Anderson and Bergmann (1951) conceived a gauge
transformation in a general field theory as ($\Phi_A$ representing
any field or field component)
$$
\delta\Phi_A = f_A \xi + f_A^\mu \partial_\mu\xi + f_A^{\mu\nu}
\partial_{\mu\nu}\xi + \dots \,,
$$
where $\xi(x^\mu)$ is an arbitrary function of the space-time
coordinates and  $\partial_\mu$ stands for the partial
derivatives. In our formulation of canonical generators, and
restricting ourselves to the language of mechanics rather than
that of field theory, this will correspond to an ansatz of the
form (see also Castellani (1982))
\begin{equation}
G(t)  = G_0 \xi(t) + G_1 \dot\xi(t) + G_2 \ddot\xi(t) + \dots =
\sum_{i=0}^{N}  G_i \xi^{(i)}(t) \,, \label{ans}
\end{equation}
with $\xi$ an arbitrary function of the time parameter and with
$G_i$ functions of the canonical variables, to be determined. We
have assumed that a finite number of terms will suffice. Let us
plug this ansatz into \bref{final4}, \bref{final5} and take into
account the arbitrariness of $\xi$. We get, from \bref{final5},
\begin{equation}
\{ G_i,\, \phi_a\} \equiv {\rm \bf pfcc}\,, \ i =
0,\dots,N.\label{check}
\end{equation}
and from \bref{final4},
\begin{eqnarray} \{ G_0,\, H'\}&\equiv&{\rm \bf
pfcc}\,,\label{G0}\\  G_{i-1} &\equiv& \{ G_i,\, H'\} + {\rm \bf
pfcc},\, \ i = 1,\dots,N.\label{iter} \\  G_N&\equiv&{\rm \bf
pfcc}\,.\label{GN}
\end{eqnarray}

The intuitive idea behind these expressions is quite clear: the
last one sets $G_N$ to be a primary first-class constraint (up to
pieces quadratic in the constraints). Next, using the iteration in
\bref{iter}, $G_{N-1}$ is found to be a secondary first-class
constraint (up to {\bf pfcc} pieces), and so on, until we reach
$G_{0}$, which is required to satisfy \bref{G0}, that puts a stop
to the stabilisation algorithm.

In addition, every $G_{i}$ must satisfy \bref{check}. It is by no
means trivial to prove that there exist solutions for the ansatz
\bref{ans}. This existence was proved in Gomis, Henneaux and Pons (1990)
under just the
conditions a) that the rank of the Hessian matrix is constant, b)
that the constraints that are initially second-class, remain
always so under the stabilisation algorithm, and c) that no
ineffective constraints\footnote{A constraint is said to be
ineffective if its differential vanishes on the constraints'
surface. See Pons, Salisbury and Shepley (2000), 
section 2, for further
considerations.} appear in the theory.

So let us suppose that these conditions are met and that a gauge
generator of the form \bref{ans} exists. This generator is made up
of first-class constraints, so it automatically satisfies the
first requirement \bref{final3}, and therefore all three
requirements \bref{final3}, \bref{final4}, \bref{final5} are
fulfilled. In addition, since it is a combination of constraints,
the value of $G$ as a conserved quantity is zero\footnote{This
assertion needs a prompt qualification in field theory because of
possible contributions from the spatial boundary in the definition
of a conserved charge from a conserved current, as happens in GR
(Regge and Teitelboim 1974, York 1972, York 1986,
Gibbons and Hawking 1977, Pons 2003).}.

\vspace{4mm}

Now let us try to recover the results of Dirac. He considered a
gauge transformation acting at time $\delta t$ and which at time
$t=0$ did not produce any change because it preserved the initial
conditions. Since
\begin{equation} \delta g (t)= \{g,\,G(t)\} =
\sum_{i=0}^{i=N} \{g,\, G_i\}\,\xi^{(i)}(t) \,,
\end{equation}
we must impose $\xi^{(i)}(0) =0, \ i = 0,\dots,N$, in order to
guarantee $\delta g (0) =0$ for generic constraints $G_i$. Then,
at an infinitesimal time $\delta t$, and to first order in $\delta
t$,
\begin{equation}
\xi^{(i)}(\delta t) =\xi^{(i)}(0) +\delta t\,\xi^{(i+1)}(0),\ i =
0,\dots,N \,,
\end{equation}
thus implying, to this order,
\begin{equation}
\xi^{(i)}(\delta t) = 0,\ i = 0,\dots,N-1 \,;\ \ \xi^{(N)}(\delta
t) =\delta t\,\xi^{(N+1)}(0) \,,
\end{equation}
where the value of $\xi^{(N+1)}(0)$ is arbitrary. The choice
$\xi^{(N+1)}(0)= v' - v$, where $v' - v$ is $v'^a - v^{a}$ as in
\bref{rule2} but for a fixed $a$ ---corresponding to a single gauge
transformation---, produces ($\epsilon=\delta t (v' - v)$, see
\bref{rule2})
\begin{equation}
\delta g (\delta t) = \epsilon \, \{g,\, G_N\}\,,
\end{equation}
which is, recalling from \bref{GN} that $G_N$ is a {\bf pfcc},
Dirac's result \bref{rule}. Here it has been obtained as a first
order computation in $ \delta t$, which was exactly Dirac's
starting point.

\vspace{4mm}

 Now we are ready to connect the concept of a gauge
transformation as a map of solutions into solutions, which is
Bergmann's view, with Dirac's concept of a gauge transformation
\emph{at a given time}, which can be understood as mapping a set
of initial conditions into a gauge equivalent set. In view of
equation \bref{ans}, it is clear that the full generator of a
gauge transformation, $G(t)$, is equivalent to a set of different,
independent generators of gauge transformations {\sl at a given
time}. It suffices to consider that, at any fixed time $t_0$, the
quantities $\xi^{(i)}(t_0)$ are just independent numerical
quantities (remember that $\xi(t)$ was an arbitrary function), and
therefore the functions $G_i$ become independent generators of
gauge transformations at a given time.

 \vspace{10mm}

\section{Comments}
Notice that if we expand to second order in $ \delta t$ the
previous expressions, the role of the secondary first-class
constraints emerges. This was probably the idea of Dirac in the
paragraph reproduced above: {\it ``For the sake of completeness,
there is a little bit or further work one ought to do which shows
that a Poisson bracket $\{H',\,\phi_a\}$ of the first-class
hamiltonian with a first-class $\phi$ is again a linear function
of first-class constraints. This can also be shown to be a
possible generator for infinitesimal contact transformation which
do not change the state"}.

It is very curious, to say the least, that such a relevant
argument, purely mathematical, does not appear more elaborated in
Dirac's book, in contrast with his previous argument, also
reproduced in section 3, where it is shown in detail that the
Poisson bracket of two primary first-class constraints must be
still a generator of gauge transformations at a given time. This
argument, which is just mentioned with a concise \emph{can also be
shown}, is crucial to understanding the origin of the
incompleteness of Dirac's analysis, because it could have been
used to show that one does not need to modify the dynamics in
order to ensure that all first-class constraints are allowed to
generate gauge transformations at a given time. At this point
Dirac seems to think of his argument as \emph{physical} rather
than \emph{mathematical}, and so he is led to believe that the
preservation, and pre-eminence, of the physical interpretation
makes it necessary to artificially modify the dynamics.

Let us rephrase our main point. Dirac attempted to carefully
distinguish the mathematical and the physical aspects of the
formulation. On the other hand, he was convinced that a physically
conceived gauge transformation at a given time, in order to be
mathematically recognised as such, should have its generator
explicitly appearing in the Hamiltonian, with its corresponding
Lagrangian multiplier. Otherwise there would have been something
wrong with the formalism. His thinking in this respect is clearly
stated in the third paragraph reproduced in section 3. Then, in
view of the fact that there {\sl could be} secondary first-class
constraints generating gauge transformations at a given time, and
that they were not present in the total Hamiltonian, he proposed
the extended Hamiltonian as the true generator of time evolution,
to prevent the formalism from any contradiction. Thus, being
afraid of an inconsistency between the mathematical formalism and
its physical interpretation, Dirac proposed to modify the dynamics
through the introduction of the Extended Hamiltonian. Considering
in his view the subsidiary role of mathematics, and in full
agreement with his approach to these disciplines (Kragh 1990), he
put physical intuition first. Regretfully he did so in this case
based on an incomplete analysis. Now, completion of his analysis
shows that there is no tension whatsoever between the mathematical
description and the physical content of gauge transformations, and
that his proposal for the Extended Hamiltonian was totally
unnecessary.

On the technical side we observe that, in constructing a gauge
transformation, Dirac looked for an infinitesimal transformation
containing an arbitrary function. The arbitrariness he found in
the functions $v$ (see \bref{rule2}), whereas the infinitesimality
he took from the time evolution, $\delta t$. This last step in his
construction is responsible for the essential incompleteness of
his analysis, because a gauge transformation \emph{at a given
time} should not have used the time as the governing parameter of
the transformation.

\vspace{4mm}

A complementary argument was also developed by Dirac. Since there
is always a mathematical jump in the process of quantization of a
classical system, he conceived that although from the classical
perspective, the classification of primary, secondary, etc.
constraints was relevant, the quantization of the system made this
nomenclature useless and for quantization the only useful concepts were those of
first-class and second-class constraints. He made his point very
clear in the last paragraph reproduced in Section 3. Obviously a
Hamiltonian including only the primary first-class constraints was
at odds with this new way of thinking, and the extended
Hamiltonian was a remedy for it. It seems likely that these
considerations helped Dirac in deciding to depart from what he
thought was deducible from the pure mathematics of the system and
to artificially introduce an ad hoc modification of the dynamics.
In Dirac's view, therefore, the argument concerning the
preservation of the physical interpretation of gauge
transformations at a given time, against the purely mathematical
interpretation (that, as we insist, he thought were mutually
inconsistent), gets somewhat mixed with arguments concerning the
process of quantization of a constrained classical system.

\vspace{4mm}

 Dirac only dealt with gauge transformations \emph{at a given time},
which can be taken as the time for the setting of the initial
conditions. But the most common view of a gauge transformation is
that of Bergmann: a {\sl symmetry} that maps \emph{entire
solutions} (or solutions defined in a region of phase space) of
the dynamics into new solutions. The misunderstandings created by
the confusion between both concepts have been enormous. Indeed
many authors take for granted from Dirac that first-class
constraints generate gauge transformations without even making the
distinction between Dirac's and Bergmann's concepts. As explained
in the last paragraph of the preceding section, the difference
between both concepts is neatly displayed when the complete
analysis of the generators of gauge transformations is performed.

Let us finally say a few words on the so-called Dirac's
conjecture. In fact, one should clearly distinguish between
Dirac's formulation of a conjecture, namely, that {\it all
first-class constraints generate gauge transformations at a given
time}, and what he thought ---wrongly in our understanding--- to be
a compulsory consequence of the assumption of this conjecture: the
necessity of modifying the dynamics. The conjecture can be proved
(Gomis, Henneaux and Pons 1990) under the same assumptions that guarantee the existence
of gauge transformations, already spelled out in Section 5.
Examples of the failure of this conjecture for some ``pathological"
models, as well as other considerations on the formalism for these models,
have been widely discussed in the literature (see for instance Allcock 1975,
Cawley 1979, Frenkel 1980, Sugano and Kimura 1983,
Costa, Girotti and Simoes 1985, Cabo and Louis-Martinez 1990,
Lusanna 1991, Lusanna 1993, Wu 1994, Miskovic and Zanelli 2003, Rothe and Rothe 2004).

\section{Saving the day: the definition of observables and the quantization
in the operator formalism}
If Dirac's approach to gauge transformations was so incomplete,
and his proposal to modify the dynamics so gratuitous and
unfounded, one may wonder to what extent has it affected the
correct development and applications of the theory. The answer is:
very little, and for various reasons. 

First, because of the
concept of observables, which will be the subject of the following
paragraph. Second, because Dirac's method of quantisation in the
operator formalism can be introduced either with the total
Hamiltonian or with the extended Hamiltonian with equivalent
results. A third reason is that important developments, for
instance the most powerful theoretical tool for the quantization
of constrained systems, the field-antifield formalism 
(Batalin and Vilkovisky 1981, Batalin and Vilkovisky 1983a,
Batalin and Vilkovisky 1983b,
see Gomis, Paris and Samuel 1995 for a general review),
did not incorporate these controversial features of Dirac's
view\footnote{In Henneaux and Teitelboim (1992) a mixed approach is
taken. These authors first assume Dirac's
analysis and consider the extended Hamiltonian formalism, on the grounds 
---see below--- that quantization in the operator formalism
only distinguishes between constraints of first or 
second class.  But the authors are able to make contact with the standard 
gauge transformations ---mapping solutions into solutions--- by considering 
the combinations of the extended Hamiltonian gauge transformations 
---generated independently by the first class constraints--- that are
compatible with setting to zero in the extended Hamiltonian the Lagrange 
multipliers associated with the secondary first-class constraints. Clearly the
total Hamiltonian formalism and its gauge transformations are
recovered. The authors consider the generalization of taking the
remaining Lagrange multipliers as new independent variables. This
approach has been extended by Garcia and Pons (2001).}. Indeed, the
natural concept of gauge transformations ---as Noether symmetries
of the action--- to be used in a path-integral framework is that of
mapping trajectories into trajectories (or field configurations
into field configurations).

Let us give some details about the first two reasons.

\vspace{5mm}

In the classical setting, an observable is defined as a function
that is gauge invariant. Consider a time dependent observable
${\mathcal O}(t)$ (the dependence with respect to the
phase space variables exists, but is not made explicit) and consider for
simplicity that there is only one gauge transformation in the
formalism, whose gauge generator is of the form \bref{ans}, that
is
$$G(t) = \sum_{i=0}^{N} G_i \xi^{(i)}(t)\,,
$$
with $G_i$ being first-class constraints. In fact they are
\emph{all} the first-class constraints of the theory if the
conditions mentioned in Section 5 are met.

Thus, ${\mathcal O}(t)$ being an observable means that the
equal-time Poisson bracket between ${\mathcal O}(t)$ and $G(t)$
vanishes:
$$
\{{\mathcal O}(t),\,G(t)\}=0\,
$$
(strictly speaking, this vanishing of the Poisson bracket needs
only to hold on the surface defined by the constraints). But,
having in mind the expansion of $G(t)$ above and the arbitrariness
of the function  $\xi(t)$, this is equivalent to the vanishing of
the Poisson bracket of ${\mathcal O}(t)$ with respect to all the
first-class constraints,
$$
\{{\mathcal O}(t),\,G_i\}=0\,, \forall\, i.
$$
The immediate consequence is that the dynamical evolution of
${\mathcal O}(t)$ is deterministic and that in this respect
\emph{it is irrelevant whatever we use the total Hamiltonian
$H_T$ or the extended Hamiltonian $H_E$}. Indeed
$$
\frac{d {\mathcal O}}{dt} = \frac{\partial {\mathcal O}}{\partial
t} + \{{\mathcal O},\,H_T\}=\frac{\partial {\mathcal O}}{\partial
t} + \{{\mathcal O},\,H_E\}\,.
$$
Since we can only attach physical significance to the gauge
invariant functions, it is clear the Dirac's proposal of the
extended Hamiltonian is harmless, as far as observables are
concerned. This key result finally saves the day for Dirac's
proposal.

\vspace{5mm}

A similar argument can be applied to Dirac's quantization in the
operator formalism, where the physical states must be gauge
invariant. Let us denote a generic Schrodinger physical state
at time $t$ by $|\psi(t)\!>$.
If the first-class constraints $G_i$ can be expressed as
linear quantum operators $\tilde G_i$, still remaining
first class, where this is understood in terms of 
the commutator of operators instead of the Poisson bracket, 
and being stable under the quantum Hamiltonian, 
then the gauge invariance property of the physical states {\it at a given time} 
becomes
\begin{equation}
\tilde G_i|\psi(t)\!>=0\,, \forall\, i\,.\label{quant}\end{equation}
So we end up with the requirement that
all first-class constraints must be enforced as operators on the
quantum state at any time on an equal footing, regardless whether they
were primary or not ---which was Dirac's idea. Second-class
constraints in Dirac
formalism are used to eliminate pairs of canonical variables
before the process of quantisation is undertaken.

Notice that equations \bref{quant} have naturally led us to
identify the gauge invariance property of the
states with the implementation of the first-class constraints as
quantum operators acting on them and giving a vanishing
eigenvalue. The difference with the classical picture is
noticeable, because the classical trajectories satisfying the
constraints are not gauge invariant and further
elimination of gauge degrees of freedom is needed, for instance through a
gauge fixing procedure. Thus the two issues of a) satisfying the constraints and 
b) being gauge invariant, which are different from the 
classical point of view, become identical in the operational quantum picture. 
This is the origin of the quantum {\it problem of time} (Isham 1992,
Kuchar 1992) for generally covariant 
systems, for which the Hamiltonian is a first class constraint.

\section{Examples}

Finally, we illustrate our discussion with two examples.

\subsection{The free relativistic particle with auxiliary variable}

The relativistic massive free particle model with auxiliary
variable is a good example for our purposes. It is described by
the Lagrangian
\begin{equation}  L = \frac{1}{2e} \dot x^\mu \dot x^\nu \eta_{\mu\nu}
    - \frac{1}{2} e m^2,
        \label{partlag}
\end{equation}
where ${ x^{\mu}}$ is the vector variable in Minkowski spacetime,
with metric $(\eta_{\mu\nu})={\rm{}diag}(-1,1,1,1)$. The parameter
$m$ is the mass and the auxiliary variable $e$ can be interpreted,
in the standard language of canonical general relativity, as a
lapse function, the world-line metric being defined by
$g_{00}=-e^2$. Its own equation of motion determines $e = (-{\dot
x}^\mu {\dot x}_\mu)^{1/2}$, and substitution of this value into
the Lagrangian\footnote{The legitimacy of this substitution is
proved in a general framework in Garcia and Pons 1997.} leads to the 
free particle
Lagrangian $L_f = -\frac{m}{2}(-{\dot x}^\mu {\dot x}_\mu)^{1/2}$.
Equation \bref{partlag} is analogous to the Polyakov Lagrangian
for the bosonic string, where the components of the world-sheet
metric are auxiliary variables. Substitution of their dynamically
determined values yields the Nambu-Goto Lagrangian, analogous to
$L_f$.

The following Noether gauge transformation is well-known to
describe the repara\-metrisation invariance for this Lagrangian
($\delta L = {d\over dt}(\epsilon L)$):
\begin{equation}
    \delta x^\mu = \epsilon {\dot x}_{\mu}, \ \delta e
    = \epsilon \dot e + {\dot \epsilon} e.
    \label{3.16}
\end{equation}
Here $\epsilon$ is an infinitesimal arbitrary function of the
evolution parameter $t$.
There is a primary constraint $\pi \approx 0$, where $\pi$ is the
variable conjugate to $e$. The only vector field in (\ref{gamma})
is now ${\bf \Gamma} = \partial/
\partial \dot e$. The condition that a function $f$ in
configuration-velocity space be projectable to phase space is
$$  {\bf\Gamma} f = \frac{\partial f}{\partial \dot e} = 0.
$$
The Noether transformation (\ref{3.16}) is not projectable to
phase space, since ${\bf\Gamma} \delta e \neq 0$.  Projectable
transformations are of the form:
\begin{equation}
    \epsilon(t,e) = {\xi(t)}/{e}.
    \label{proj3.16}
\end{equation}
The Noether variations then become:
\begin{equation}
    \delta x^{\mu} = \xi \frac{{\dot x}^\mu}{e},\ \delta e = \dot \xi.
    \label{del-x}
\end{equation}
The arbitrary function describing the Noether gauge transformation
is now $\xi(t)$.

The canonical Hamiltonian is
$$  H = \frac{1}{2} e ( p^\mu p_\mu + m^2),
$$
where $p_\mu$ is the variable canonically conjugate to $x^\mu$.
The evolution operator vector field $\{-,H\}+ \lambda(t)
\{-,\pi\}$ yields the secondary constraint $\frac{1}{2} (p^\mu
p_\mu + m^2) \approx 0$. Both the primary and the secondary
constraints are first class. The arbitrary function $\lambda$ is a
reflection of the gauge invariance of the model. The solutions of
the equations of motion are:
$$  x^\mu(t) = x^\mu(0) + p^\mu(0) \left(e(0) t
    + \int_0^t d\tau \int_0^\tau d\tau' \, \lambda(\tau')\right),
$$
$$  e(t) = e(0) + \int_0^t d\tau \, \lambda(\tau),
$$
$$  p^\mu(t) = p^\mu(0),
$$
$$  \pi(t) = \pi(0),
$$
with the initial conditions satisfying the constraints.

The canonical generator of gauge transformations, satisfying
\bref{final3}, \bref{final4} and \bref{final5}, is
\begin{equation}
G = \xi(t)\frac{1}{2} (p^\mu p_\mu + m^2) + \dot\xi(t)
\pi\,.\label{ggfree}
\end{equation}

Gauge transformations relate trajectories obtained through different
choices of $\lambda(t)$.  Consider an infinitesimal change
$\lambda \rightarrow \lambda + \delta\lambda$.  Then the change
in the trajectories (keeping the initial conditions intact) is:
$$  \delta x^\mu(t) = p^\mu(0)
    \left(\int_0^t d\tau \int_0^\tau d\tau' \,
    \delta\lambda(\tau')\right),
$$
$$  \delta e(t) = \int_0^t d\tau \, \delta\lambda(\tau),
$$
$$  \delta p^\mu(t) = 0 , \, \delta \pi(t) = 0,
$$
which is nothing but a particular case of the projectable gauge
transformations displayed above with
$$\xi(t) =\int_0^t d\tau \int_0^\tau d\tau' \, \delta\lambda(\tau').
$$

Notice that the structure of the gauge generator \bref{ggfree} is
that of \bref{ans}. It is only this particular combination of
primary and secondary first-class constraints that generates gauge
transformations mapping solutions to solutions.

\subsection{Maxwell theory}

The case of pure electromagnetism is described with the Lagrangian
$${\mathcal L}_M = -\frac{1}{4} F_{\mu \nu}F^{\mu \nu} \,,
$$
where $F_{\mu \nu}= \partial_\mu A_\nu - \partial_\nu A_\mu$ and
$A_\mu$ is the Maxwell gauge field. We take again the metric in
Minkowski spacetime as $(\eta_{\mu\nu})={\rm{}diag}(-1,1,1,1)$.
The canonical Hamiltonian is
$$H_c = \int d{\bf x} \left[\frac{1}{2}({\vec \pi}^2
+ {\vec B}^2) + {\vec \pi} \cdot \nabla A_0\right] \,,
$$
where the electric field ${\vec \pi}$ stands for the spatial
components of $\pi^\mu$, the variables canonically conjugate to
$A_\mu$. The Lagrangian definition of $\pi^\mu$ is $\hat\pi^\mu =
-F^{0\mu}$ and so $\pi^0$ is a primary constraint, $\pi^0 \approx
0$. The magnetic field is defined as $B_i=
\frac{1}{2}\epsilon_{ijk}F^{jk}$. Stability of the constraint
$\pi^0$ under the Hamiltonian dynamics leads to the secondary
constraint $\dot\pi^0 = \{\pi^0, H_c\} = \nabla \cdot {\vec \pi}
\approx 0$. Both constraints are first-class and no more
constraints arise.

Now, similarly to the previous example, the gauge generator takes
the form
$$
G[t] = \int d^3\!x \, \left[-\dot{\Lambda}({\bf x}, t)\,
\pi^0({\bf x}, t) + \Lambda({\bf x},t)\, \nabla \!\cdot\!{\vec
\pi}({\bf x}, t)\right] \,
$$
with ${\Lambda}({\bf x}, t)$ an arbitrary scalar function of the
space-time coordinates. The gauge transformation of the gauge
field is then
$$\delta A_{\mu} =
\{A_{\mu}, \, G\} = -\partial_\mu{\Lambda} \,,
$$
which is the usual Noether $U(1)$ symmetry for the Lagrangian
${\mathcal L}_M$. Let us observe again that a primary and a
secondary constraint are necessary to build the gauge generator.
Notice also that the particular combination of both constraints,
together with the role of the function ${\Lambda}$ and its time
derivative, eventually ensures that the gauge field $A_{\mu}$
transforms covariantly.

\section{Acknowledgments}
I thank Jeremy Butterfield for encouragement and many valuable
suggestions on the manuscript. I also thank Lu\'\i s Navarro for
his useful comments. This work is partially supported by MCYT FPA
2001-3598 and CIRIT GC 2001SGR-00065.
\section{References}

\hspace{4mm}
R.~Abraham and J.~E.~Marsden (1978), 
{\em Foundations of mechanics}, 2nd
ed. Benjamin/Cummings Publishing Co. Reading, Mass.\vspace{2mm}

J.~L.~Anderson and P.~G.~Bergmann (1951),
``Constraints In Covariant Field Theories,''
{\em Physical\ Review, 83}, 1018-1025.\vspace{2mm}

G.~R.~Allcock (1975), ``The intrinsic properties of rank and nullity of 
the Lagrange bracket in the one dimensional calculus of variations,''  
{\em Philosophical\ Transactions\ of\
the\ Royal\ Society\ of\ London\ A, 279}, 487-545. \vspace{2mm}

R. ~Arnowitt, S. ~Deser, and C.~W.~Misner (1962),
 ``The dynamics of general relativity,''
in {\it Gravitation: An Introduction to Current Research}.
Edited by L.\ Witten
(John Wiley \& Sons, New York, 1962), 227--265.\vspace{2mm}

R.~Banerjee, H.~J.~Rothe and K.~D.~Rothe (1999),
``Hamiltonian approach to Lagrangian gauge symmetries,''
{\em Physics\ Letters\ B, 463}, 248-251
[arXiv:hep-th/9906072].\vspace{2mm}

I.~A.~Batalin and G.~A.~Vilkovisky (1981),
``Gauge Algebra And Quantization,''
{\em Physics\ Letters\ B, 102}, 27-31. \vspace{2mm}

I.~A.~Batalin and G.~A.~Vilkovisky  (1983a),
``Feynman Rules For Reducible Gauge Theories,''
{\em Physics\ Letters\ B, 120}, 166-170. \vspace{2mm}

I.~A.~Batalin and G.~A.~Vilkovisky (1983b),
``Quantization Of Gauge Theories With
Linearly Dependent Generators,'' {\em Physical\ Review\  D, 28} 2567-2582
[Erratum (1984) {\em Physical\ Review\   D, 30}, 508].\vspace{2mm}

C.~Batlle, J.~Gomis, J.~M.~Pons and N.~Roman (1986),
``Equivalence Between The Lagrangian And Hamiltonian Formalism For Constrained
Systems,''
{\em Journal\ of\ Mathematical\ Physics, 27}, 2953.\vspace{2mm}

C.~Batlle, J.~Gomis, X.~Gracia and J.~M.~Pons (1989), ``Noether's Theorem
And Gauge Transformations: Application To The Bosonic String And
$Cp^(2N-1)$ Model,'' {\em Journal\ of\ Mathematical\ Physics, 30}, 1345-1350.
\vspace{2mm}

P.~G.~Bergmann (1949),
``Non-Linear Field Theories,'' 
{\em Physical\ Review, 75}, 680-685.\vspace{2mm}

P.~G.~Bergmann and J. ~H. ~M. ~Brunings (1949),
`` Non-Linear Field Theories II. Canonical Equations and Quantization,''
{\em Reviews\ of\ Modern\ Physics, 21}, 480-487.\vspace{2mm}

J.~D.~Bjorken and S.~D.~Drell (1965), 
{\em Relativistic Quantum fields}, New
York, McGraw-Hill.\vspace{2mm}

A.~Cabo and D.~Louis-Martinez (1990),
``On Dirac's Conjecture For Hamiltonian Systems With First And Second Class
Constraints,''
{\em Physical\ Review\  D, 42}, 2726-2735. \vspace{2mm}

L.~Castellani (1982),
``Symmetries In Constrained Hamiltonian Systems,''
{\em Annals\ of\ Physics, 143}, 357-371.\vspace{2mm}

R.~Cawley (1979), ``Determination of the Hamiltonian in the Presence of Constraints,''
{\em Physical\ Review\ Letters, 42}, 413-416. \vspace{2mm}

M.~E.~V.~Costa, H.~O.~Girotti and
T.~J.~M.~Simoes (1985), ``Dynamics Of Gauge Systems And Dirac's Conjecture,''
{\em Physical\ Review\ D, 32}, 405-410. \vspace{2mm}

B.~S.~DeWitt (1963), in {\em Relativity, Groups and Topology},
Gordon and Breach, New York.\vspace{2mm}

P.~A.~M.~Dirac (1949), ``Forms of relativistic dynamics,'' 
{\em Reviews\ of\ Modern\ Physics, 21},
 392-399.\vspace{2mm}

P.~A.~M.~Dirac (1950), ``Generalized Hamiltonian Dynamics,'' 
{\em Canadian\ Journal\ of\ Mathematics, 2}, 129-148.\vspace{2mm}

P.~A.~M.~Dirac (1951), ``The Hamiltonian form of field dynamics,'' 
{\em Canadian\ Journal\ of\ Mathematics, 3}, 1-23.\vspace{2mm}

P.~A.~M.~Dirac (1958),
``The Theory Of Gravitation In Hamiltonian Form,''
{\em Proceedings\ of\ the\ Royal\ Society\ of\ London\ A, 246}, 333-343.\vspace{2mm}

P.~A.~M.~Dirac (1959), ``Fixation Of Coordinates In The Hamiltonian
Theory Of Gravitation,'' {\em Physical\ Review, 114}, 924-930.\vspace{2mm}

P.~A.~M.~Dirac (1964),
 {\em Lectures on Quantum Mechanics},
    Yeshiva\ University\ Press\ , New York.\vspace{2mm}

J.~Earman (2003),
``Tracking down gauge: An ode to the constrained Hamiltonian formalism,''
in K. ~Brading and E. ~Castellani (eds.),
{\em Symmetries in Physics: Philosophical
   Reflections}, Cambridge University Press.\vspace{2mm}

L.~D.~Faddeev and R.~Jackiw (1988), ``Hamiltonian Reduction Of
Unconstrained And Constrained Systems,'' {\em Physical\ Review\ Letters\,
60}, 1692-1694.\vspace{2mm}

E.~Fermi (1932),
``Quantum theory of radiation,'' {\em Reviews\ of\ Modern\ Physics, 4},
87-132.\vspace{2mm}

R.~P.~Feynman (1948),
``Space-Time Approach To Nonrelativistic Quantum Mechanics,''
   {\em Reviews\ of\ Modern\ Physics, 20}, 367-387.\vspace{2mm}

A.~Frenkel (1980),
``Comment On Cawley's Counterexample To A Conjecture Of Dirac,''
{\em Physical\ Review\ D, 21}, 2986-2990. \vspace{2mm}

J.~A.~Garcia and J.~M.~Pons (1997),
``Equivalence of Faddeev-Jackiw and Dirac approaches for gauge theories,''
{\em International\ Journal\ of\ Modern\ Physics\ A,  12}, 451-464
[arXiv:hep-th/9610067].\vspace{2mm}

J.~A.~Garcia and J.~M.~Pons (2000),
``Rigid and gauge Noether symmetries for constrained systems,''
{\em International\ Journal\ of\ Modern\ Physics\ A, 15}, 4681-4721
[arXiv:hep-th/9908151].\vspace{2mm}

J.~A.~Garcia and J.~M.~Pons (2001),
``Lagrangian Noether symmetries as canonical transformations,''
{\em International\ Journal\ of\ Modern\ Physics\ A,  16}, 3897-3914
[arXiv:hep-th/0012094].\vspace{2mm}

G.~W.~Gibbons and S.~W.~Hawking (1977),
``Action Integrals And Partition
Functions In Quantum Gravity,'' {\em Physical\ Review\ D,  15},
2752-2756.\vspace{2mm}

J.~Gomis, M.~Henneaux and J.~M.~Pons (1990), 
``Existence Theorem For
Gauge Symmetries In Hamiltonian Constrained Systems,'' {\em Classical\ and\
Quantum\ Gravity, 7}, 1089-1096.\vspace{2mm}

J.~Gomis, J.~Paris and S.~Samuel (1995),
``Antibracket, antifields and gauge theory quantization,''
{\em Physics\ Reports, 259}, 1-145
[arXiv:hep-th/9412228].\vspace{2mm}

X.~Gr\`acia and J.~M.~Pons (1988),
``Gauge Generators, Dirac's Conjecture And Degrees Of Freedom For Constrained
Systems,''
{\em Annals\ of\ Physics, 187}, (1988) 355-368.\vspace{2mm}

X.~Gr\`acia and J.~M.~Pons (2000), 
``Canonical Noether symmetries and
commutativity properties for gauge systems,'' {\em Journal\ of\ 
Mathematical\ Physics, 41}, 7333-7351 [arXiv:math-ph/0007037].\vspace{2mm}

X.~Gr\`acia and J.~M.~Pons (2001), 
``Singular lagrangians: some geometric
structures along the Legendre map,''{\em  Journal\ of\ Physics\ A, 34},
3047-3070 [arXiv:math-ph/0009038].\vspace{2mm}

M.~Henneaux and C.~Teitelboim (1992),
{\em Quantization of gauge systems}, Princeton University Press.\vspace{2mm}

C.~J.~Isham (1992),
``Canonical quantum gravity and the problem of time,'' in L. A. Ibort 
and M. A. Rodriguez (eds.), {\em Integrable systems, quantum groups, and 
quantum field theories}, pp 157-287. Boston, Kluwer Academic.
[arXiv:gr-qc/9210011]\vspace{2mm}

H.~Kragh (1990), {\em Dirac, a scientific biography}, Cambridge University
Press.\vspace{2mm}

K.~V.~Kuchar (1992),
``Time and interpretations of quantum gravity,'' in G. Kunstatter, D. Vincent and
J. Williams (eds.), {\em Proceedings of the 4th canadian conference on general 
relativity and relativistic astrophysics}, pp 211-314.\vspace{2mm}

J.~Lee and R.~M.~Wald (1990), ``Local Symmetries And Constraints,'' 
{\em Journal\ of\ Mathematical\ Physics, 31}, 725-743.\vspace{2mm}

L.~Lusanna (1991), ``The second Noether theorem as the basis of the theory of
singular  Lagrangians and Hamiltonian constraints'',
                 {\em  La Rivista del Nuovo Cimento, vol. 14, n.3} \vspace{2mm}

L.~Lusanna (1993), ``The Shanmugadhasan canonical transformation, function groups
and the extended second Noether theorem'',
{\em  International\ Journal\ of\ Modern\ Physics, A8}, 4193-4233.\vspace{2mm}

L.~Lusanna and M.~Pauri (2003),
``General Covariance and the Objectivity of Space-Time Point-Events: The
Physical Role of Gravitational and Gauge Degrees of Freedom in General
Relativity,''
[arXiv:gr-qc/0301040].\vspace{2mm}

O.~Miskovic and J.~Zanelli (2003),
``Dynamical structure of irre\-gular constrained systems,''
{\em  Journal\ of\ Mathematical\ Physics, 44}, 3876-3887.
[arXiv:hep-th/0302033]. \vspace{2mm}

J.~M.~Pons and D.~C.~Salisbury,
``The issue of time in generally covariant theories and the Komar-Bergmann
approach to observables in general relativity,''
  [arXiv:gr-qc/0503013].\vspace{2mm}

J.~M.~Pons, D.~C.~Salisbury and L.~C.~Shepley (1999), ``Reduced phase
space: Quotienting procedure for gauge theories,'' {\em  Journal\ of\ Physics\ A,
 32}, 419-430 [arXiv:math-ph/9811029].\vspace{2mm}

J.~M.~Pons, D.~C.~Salisbury and L.~C.~Shepley (2000),
``Gauge group and reality conditions in Ashtekar's complex formulation of
canonical gravity,''
{\em  Physical\ Review\ D,  62}, 064026 (15 pp.)
[arXiv:gr-qc/9912085].\vspace{2mm}

J.~M.~Pons (2003), ``Boundary conditions from boundary terms, Noether
charges and the trace  K Lagrangian in general relativity,'' {\em  General\
Relativity\ and\ Gravitation, 35}, 147-174  [arXiv:gr-qc/0105032].
\vspace{2mm}

T.~Regge and C.~Teitelboim (1974), ``Role Of Surface Integrals In The
Hamiltonian Formulation Of General Relativity,'' {\em  Annals\ of\ Physics,
 88}, 286-318.\vspace{2mm}

H.~J.~Rothe and K.~D.~Rothe (2004),
``Gauge identities and the Dirac conjecture,''
[arXiv:hep-th/0406044].\vspace{2mm}

J.~Sniatycki (1974), ``Dirac brackets in geometric dynamics,''{\em   Annales\  
de\ l'Institut\ Henri\ Poincar\'e\ A, 20}, 365-372.\vspace{2mm}

R.~Sugano and T.~Kimura (1983),
``Pathological Dynamical System With Constraints,''
{\em  Progress\ of\ Theoretical\ Physics, 69}, 1241-1255. \vspace{2mm}

J.~W.~York, (1972)
``Role Of Conformal Three Geometry In The Dynamics Of Gravitation,''
{\em  Physical\ Review\ Letters, 28}, 1082-1085.\vspace{2mm}

J.~W.~York (1986),
``Boundary Terms in the Action Principles of General Relativity,''
{\em  Foundations\ of\ Physics, 16}, 249-257.\vspace{2mm}

S.~Weinberg (1995),{\em The Quantum Theory Of Fields}. Vol. 1: Foundations,
chap. 9. Cambridge University Press.\vspace{2mm}

B.~C.~Wu (1994),
``Dirac's Conjecture,''
{\em  International\ Journal\ of\ Modern\ Physics, 33}, 1529-1533. \vspace{2mm}


\end{document}